\documentclass[aps,pre,superscriptaddress,bibtex,twocolumn,longbibliography]{revtex4-1}

\usepackage[utf8]{inputenc}
\usepackage[T1]{fontenc}

\usepackage[tbtags]{amsmath}
\usepackage{amssymb}
\usepackage{mathtools}

\usepackage{float}

\usepackage[section]{placeins}
\usepackage[margin=1in]{geometry}

\usepackage[textsize=scriptsize,]{todonotes}
\makeatletter \def\@captype{figure} \makeatother
\usepackage{xcolor}

\usepackage[colorlinks=true,urlcolor=blue,linkcolor=blue,citecolor=blue]{hyperref}

\newcommand{\dd}{\mathrm{d}}
\newcommand{\avg}[1]{\langle #1 \rangle}

\newcommand{\abs}[1]{\lvert #1 \rvert}

\newcommand{\occup}{n}

\newcommand{\genmatDrate}{\genmatDrate_{\Doobletter}}

\newcommand{\npartcl}{N}

\newcommand{\Eorder}{m}
\newcommand{\dens}{\rho}
\newcommand{\densxt}{\dens(\pos,\tv)}
\newcommand{\meandens}{\rho_0}

\newcommand{\tv}{t}
\newcommand{\pos}{x}
\newcommand{\diffcoef}{D}
\newcommand{\mobility}{\sigma}
\newcommand{\E}{E}
\newcommand{\Ed}{\epsilon} 
 
\newcommand{\Epampl}{\lambda}

\newcommand{\lattsize}{L}

\newcommand{\opar}{z} 
\newcommand{\oparm}{\opar_{\Eorder}}

\newcommand{\wnumber}{2\pi}
\newcommand{\binder}{U_4}

\newcommand{\clengthexp}{\nu}
\newcommand{\oparexp}{\beta}
\newcommand{\suscepexp}{\gamma}

\begin{document}

\title{Critical behavior of a programmable time-crystal lattice gas}
\author{R. Hurtado-Guti\'errez}
\affiliation{Departamento de Electromagnetismo y F\'isica de la Materia, Universidad de Granada, Granada 18071, Spain}
\affiliation{Institute Carlos I for Theoretical and Computational Physics, Universidad de Granada, Granada 18071, Spain}
\author{C. P\'erez-Espigares}
\affiliation{Departamento de Electromagnetismo y F\'isica de la Materia, Universidad de Granada, Granada 18071, Spain}
\affiliation{Institute Carlos I for Theoretical and Computational Physics, Universidad de Granada, Granada 18071, Spain}
\author{P.I. Hurtado}
\affiliation{Departamento de Electromagnetismo y F\'isica de la Materia, Universidad de Granada, Granada 18071, Spain}
\affiliation{Institute Carlos I for Theoretical and Computational Physics, Universidad de Granada, Granada 18071, Spain}
\date{\today}

\begin{abstract}
Time crystals are many-body systems whose ground state spontaneously breaks time-translation symmetry and thus exhibits long-range spatiotemporal order and robust periodic motion. Using hydrodynamics, we have recently shown how an $m$th-order external packing field coupled to density fluctuations in driven diffusive fluids can induce the spontaneous emergence of time-crystalline order in the form of $m$ rotating condensates, which can be further controlled and modulated. Here we analyze this phenomenon at the microscopic level in a paradigmatic model of particle diffusion under exclusion interactions, a generalization of the weakly asymmetric simple exclusion process with a configuration-dependent field called the time-crystal lattice gas. Using extensive Monte Carlo simulations, we characterize the nonequilibrium phase transition to these complex time-crystal phases for different values of $m$, including the order parameter, the susceptibility and the Binder cumulant, from which we measure the critical exponents, which turn out to be within the Kuramoto universality class for oscillator synchronization. We also elucidate the condensates density profiles and velocities, confirming along the way a scaling property predicted for the higher-order condensate shapes in terms of first-order ones, discussing also novel possibilities for this promising route to time crystals.
\end{abstract}

\maketitle

\section{Introduction}

A time crystal \cite{wilczek12a, shapere12a} is a many-body system where time-translation symmetry is spontaneously broken. This leads to a continuous and robust oscillatory behavior over time in the ground state \cite{zakrzewski12a,richerme17a,yao18a,sacha18a,sacha20a}. While breaking symmetries is common in nature, as seen in many spontaneous symmetry-breaking phenomena \cite{binney92a}, time-translation symmetry was once thought to be unbreakable. However, recent progress has shown that both continuous and discrete time-translation symmetries can indeed be broken, resulting in continuous and discrete time crystals, respectively.

In quantum systems, continuous time crystals are forbidden in equilibrium short-range systems due to various no-go theorems \cite{bruno13a,nozieres13a,watanabe15a,kozin19a}. These limitations are overcome in nonequilibrium dissipative settings, which allow for the formation of continuous time crystals \cite{iemini18a,buca19a,kessler19a,carollo20a,carollo22a}. Such a continuous time crystal was recently observed experimentally in an atom-cavity system \cite{kongkhambut22a}. Discrete time crystals, on the other hand, can appear as a subharmonic response to periodic (Floquet) driving. These have been theoretically predicted \cite{else16a, moessner17a, yao17a, gong18a, gambetta19a, khemani19a, pizzi19a, lazarides20a, else20a, pizzi21a, yousefjani25a} and experimentally observed in isolated \cite{zhang17a, choi17a, rovny18a, smits18a, autti18a, osullivan20a, kyprianidis21a, randall21a, xiao22a} and dissipative \cite{kessler21a, kongkhambut21a} systems.

Classical systems can also exhibit time-crystalline order \cite{pizzi21b, pizzi21c}. For instance, discrete time crystal phases have been predicted in a periodically driven two-dimensional ($2d$) Ising model \cite{gambetta19b} and in a one-dimensional ($1d$) system of coupled nonlinear pendula at finite temperature \cite{yao18b}, with experimental demonstrations in a classical network of dissipative parametric resonators \cite{heugel19a}. A continuous time crystal has been observed in a classical $2d$ array of plasmonic metamolecules, displaying superradiant-like transmissivity oscillations \cite{liu23a}. Despite these advances, a comprehensive approach to creating programmable time-crystal phases remains elusive.
    
In a recent work inspired by the rare event statistics of some driven diffusive systems \cite{hurtado-gutierrez20a,hurtado-gutierrez23a}, we uncovered a novel mechanism to build  time-crystal phases based on the concept of \emph{packing field}. Indeed, when conditioned to sustain a time-averaged particle current well below its typical value, some interacting particle systems exhibit a dynamical phase transition to a traveling wave phase \cite{bertini05a, bertini06a, derrida07a, perez-espigares13a, hurtado14a, lazarescu15a, shpielberg18a, perez-espigares19a}. The distinguishing feature of this nonequilibrium phase is the emergence of a particle condensate that moves in a periodic fashion, hindering the overall particles' motion and thus increasing the probability of the associated current fluctuation. It can be shown that this phase shares the properties of a time crystal, as it displays robust, coherent periodic motion and long-range spatiotemporal order despite the stochasticity of the underlying dynamics, thus breaking the continuous time-translation symmetry present in the system \cite{hurtado-gutierrez20a,hurtado-gutierrez23a}. 
    
\begin{figure}[h!]
\includegraphics[width=\linewidth]{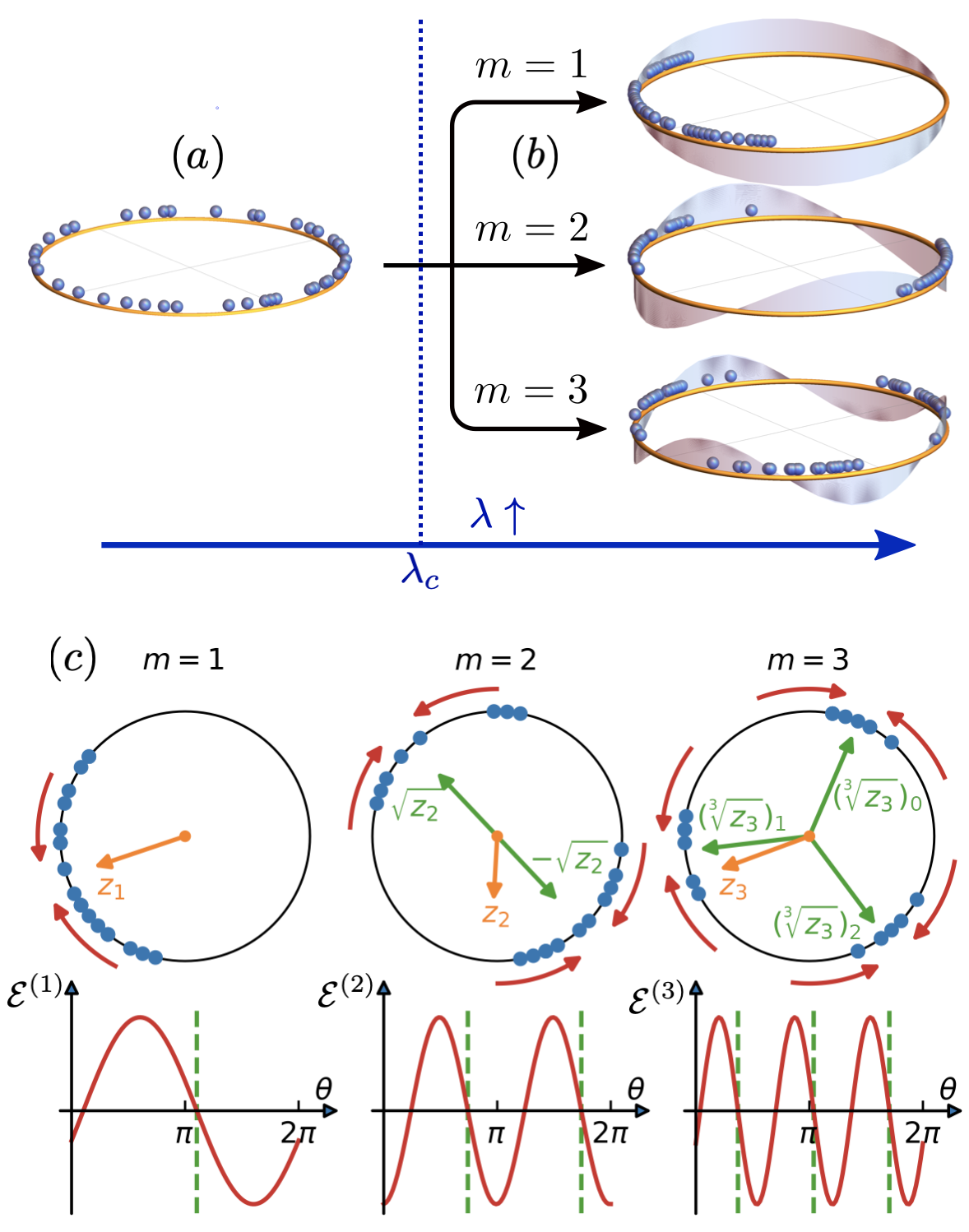}
\caption{\textbf{The packing-field mechanism.} 
    (a) A stochastic particle fluid in the presence of a constant driving field $\epsilon$ sustains a net current of particles with a homogeneous density structure on average subject to small density fluctuations. (b) By switching on a $\Eorder$-th-order packing field $\mathcal{E}^{(m)}$ with a coupling $\lambda$ beyond a critical value  $\lambda_c$, an instability is triggered to a complex time-crystal phase characterized by the emergence of $\Eorder$ rotating particle consensates with a velocity controlled by the driving field. (c) Sketch showing the working of the packing field $\mathcal{E}^{(m)}$ for different packing orders $\Eorder = 1,2,3$. The top row shows possible condensate configurations for varying $m$, together with the corresponding packing order parameter $\oparm = \abs{\oparm}e^{i\varphi_m}$ (orange vector). Its magnitude $\abs{\oparm}$ measures the amount of particle packing around the $m$ emergent localization centers located at angular positions $\phi_m^{(j)}=(\varphi_m + 2\pi j)/m$ (see green arrows), with $j\in[0,m-1]$, i.e. the arguments of $(\sqrt[m]{\opar_m})_j = \sqrt[m]{\opar_m}\textrm{e}^{\textrm{i}2\pi j/\Eorder}$. The red, external arrows in top panels signal the local direction of the packing field, which hinders particle motion ahead of each localization point and pushes particles lagging behind, promoting in this way particle packing. The bottom panels display the value of the packing field as a function of the angular position along the ring, while the green dashed lines mark the emergent localization centers at $\phi_m^{(j)}$. 
}
\label{fig1}
\end{figure}

Interestingly, these rare fluctuations can be made typical using the Doob's transform as a tool \cite{doob57a, chetrite15a, chetrite15b, bertini15a, carollo18b, simon09a, jack10a, popkov10a}, which provides the physical dynamics responsible for a given fluctuation. A detailed analysis of such underlying dynamics shows that the mechanism which triggers the instability leading to the time crystal dynamical phase can be interpreted as an external packing field that pushes particles that lag behind the emergent condensate's center of mass while restraining those moving ahead, see left column in Fig.~\ref{fig1}.(c). This amplifies naturally-occurring fluctuations of the particles' spatial packing, a nonlinear feedback mechanism that eventually leads to a time-crystal. These ideas motivated the introduction of a novel model, the time-crystal lattice gas (TCLG) \cite{hurtado-gutierrez20a}, which implements a controllable external packing field and exhibits the fundamental properties of the previous dynamical phase transition in its steady state behavior. In particular, the TCLG displays a (steady-state) continuous phase transition to a time-crystal phase for a sufficiently strong packing field.

Mathematically, the action of the packing field can be seen as a controlled excitation of the first Fourier mode of the density field around the instantaneous center-of-mass position \cite{hurtado-gutierrez20a}. A natural question then concerns the system response to the excitation of higher, $m$th-order modes (with $m>1$). We have recently demonstrated using hydrodynamics and a local stability analysis \cite{hurtado-gutierrez25a} that this excitation mechanism opens the door to  fully programmable continuous time-crystal phases in driven diffusive fluids, characterized by an arbitrary number $m$ of rotating condensates, see Fig.~\ref{fig1}, which can be further enhanced with higher-order modes. In particular, hydrodynamics allows the derivation of general properties of the condensates density profiles and velocities, as well as a scaling property of higher-order traveling condensates in terms of first-order ones. These findings were illustrated in \cite{hurtado-gutierrez25a} for several paradigmatic driven diffusive systems \cite{spohn12a, kipnis82a, spitzer70a, derrida98a, katz84a, hager01a, krapivsky13a}, where programmable time-crystal phases were demonstrated and characterized in all cases, finding along the way a novel explosive time-crystal phase transition, i.e., a discontinuos transition to a time-crystal phase, for certain nonlinear transport coefficients. Programmable in this context means that we can control on demand the number, shape, and velocity of the emerging condensates. In this way, these results demonstrate the utility and rich possibilities of this promising route to time crystals, which may serve as a robust platform to engineer these complex nonequilibrium phases of matter and study how to exploit them at the technological level. 

In this work we analyze this phenomenon at the microscopic level in a paradigmatic model of particle diffusion under exclusion interactions, a generalization of the weakly asymmetric simple exclusion process \cite{spitzer70a,derrida98a} with a configuration-dependent field called the time-crystal lattice gas (TCLG) \cite{hurtado-gutierrez20a}. Using extensive Monte Carlo simulations, we characterize the nonequilibrium phase transition to these complex time-crystal phases for different values of the packing-field order $m$. In particular, we measure the order parameter in each case, the susceptibility and the Binder cumulant, comparing with the hydrodynamic predictions whenever possible. We also determine the critical exponents for these time-crystal phase transitions, which turn out to be within the Kuramoto universality class that characterizes the synchronization of oscillators, independently of the packing order $m$. We also characterize the condensates density profiles and velocities, confirming the mentioned scaling property predicted for the higher-order condensate shapes in terms of first-order ones.

The paper is structured as follows. In Sec.~\ref{sec:model} we define the TCLG model, discussing in detail the workings of the generalized packing field mechanism, introduced in \cite{hurtado-gutierrez25a} at the hydrodynamic level. We also discuss in this section the hydrodynamic equations describing this model and their predictions, including the local stability of the homogeneous solution which can be used to determine the critical threshold for the time-crystal phase transition to occur. Sec.~\ref{sec:numerics} is devoted to the microscopic characterization of the observed phase transition via extensive Monte Carlo simulations of the TCLG model for different packing orders $m$. In particular we fully validate the hydrodynamic predictions \cite{hurtado-gutierrez25a}, characterizing the underlying phase transitions and their universality class. We also analyze the dependence of the condensates' density profiles on the model parameters, as well as their velocities. In addition, we exemplify the versatility of the packing-field route to engineer and control complex time-crystal phases by showing possible time-dependent modulated protocols that allow mixing and coexistence of multiple time-crystalline matter waves. Finally, we end the paper with a discussion of the results and the possible new avenues of research to exploit the packing-field route to time crystals.

\section{The packing-field route to time crystals}
\label{sec:model}
\newcommand{\kp}{k^{\prime}}
    
We hence consider a stochastic lattice gas with $\npartcl$ particles moving on a one-dimensional ($1d$) periodic ring with $\lattsize\ge\npartcl$ lattice sites, so that the global density is $\rho_0\equiv N/L\le 1$. Each lattice site $k\in[1,L]$ might be either empty (occupation number $n_k=0$) or occupied by one particle at most ($n_k=1$), defining a system configuration $\mathbf{n}\equiv \{n_k, k\in[1,L]\}$. A particle at site $k$ can jump randomly to a empty neighboring site in the clockwise ($-$) or anti-clockwise ($+$) direction with rates $p^{\pm}_{k}(\mathbf{n}) = \frac{1}{2}\, \mathrm{e}^{\pm E_k(\mathbf{n}) / \lattsize}$, with $\E_k(\mathbf{n})$ an external field that may depend on the particular site $k$ and the corresponding system configuration $\mathbf{n}$. Following the ideas introduced in \cite{hurtado-gutierrez20a,hurtado-gutierrez25a} and discussed above, the TCLG is characterized by an external field with two components, $\E_k(\mathbf{n}) = \epsilon + \lambda \mathcal{E}^{(m)}_k(\mathbf{n})$. Here $\epsilon$ is a constant driving field pushing the particles homogeneously in a given direction, while $\lambda\ge 0$ is the coupling to an $m$th-order packing field $\mathcal{E}^{(m)}_k(\mathbf{n})$ defined as
\begin{equation}
\mathcal{E}^{(m)}_k(\mathbf{n}) = \abs{\oparm} \sin(\varphi_m - \frac{2\pi \Eorder k}{\lattsize}) \, ,
\label{Epack}
\end{equation}
with $2\pi k/L=\theta_k$ being the particles' angular positions, and $\oparm(\mathbf{n})$ the complex $\Eorder$th-order packing order parameter, also known as Kuramoto-Daido parameter in synchronization literature \cite{kuramoto84a,kuramoto87a,daido92a,pikovsky03a,acebron05a}, 
\begin{equation}
\oparm(\mathbf{n}) = \frac{1}{\npartcl}{\sum_{\kp=1}^\lattsize \occup_{\kp} e^{i 2\pi \Eorder \kp/\lattsize}} \equiv \abs{\oparm} \textrm{e}^{\textrm{i}\varphi_m} \, ,
\label{zparam}
\end{equation}
with magnitude $\abs{\oparm}$ and argument $\varphi_m$, which sets the angular position of the first condensate (given by $\varphi_m/m$), as we shall explain below.  The constant driving $\epsilon$ gives rise to a net particle current in the desired direction, and as we will see below, controls the velocity of the resulting particle condensates and the asymmetry of the associated density waves. On the other hand, the packing field $\mathcal{E}^{(m)}_k(\mathbf{n})$ pushes particles locally towards $\Eorder$ \emph{emergent localization centers} where particles are most clustered, with an amplitude proportional to the instantaneous magnitude of the packing order parameter $\abs{\oparm}$ and the coupling constant $\lambda$. The angular position along the $1d$ ring of the $m$ emergent localization centers is given by $\phi_m^{(j)}=(\varphi_m + 2\pi j)/m$, with $j\in[0,m-1]$, i.e. the arguments of the $m$ roots of the complex $m$th-order packing order parameter, $(\sqrt[m]{\opar_m})_j = \sqrt[m]{\opar_m}\textrm{e}^{\textrm{i}2\pi j/\Eorder}$, as illustrated in Fig.~\ref{fig1}.(c).
   
The packing field mechanism works by restraining the motion of particles ahead of the closest localization center, i.e. for nearby lattice sites $k$ where $\mathcal{E}^{(m)}_k(\mathbf{n})<0$, see Eq. \eqref{Epack}, while pushing particles lagging behind this point (where $\mathcal{E}^{(m)}_k(\mathbf{n})>0$), see Fig.~\ref{fig1}. The strength of the packing mechanism is proportional to the coupling constant $\Epampl$ and the magnitude $\abs{\oparm}$ of the packing order parameter, which measures the concentration of particles around the emergent localization centers. For large enough values of $\Epampl$, the packing field leads to a nonlinear feedback mechanism which amplifies the density fluctuations naturally present in the system, resulting eventually in a phase transition to a time-crystal phase, exhibiting the fingerprints of spontaneous time-translation symmetry breaking \cite{hurtado-gutierrez20a,hurtado-gutierrez25a}. Interestingly, the packing field can also be expressed as a long-range pairwise interaction between all particles. Indeed, by expanding the value of $\oparm$, and using some basic trigonometric identities, it is easy to show that \cite{hurtado-gutierrez20a,hurtado-gutierrez25a}
\begin{equation}
\label{eq:extfield_pairs}
\mathcal{E}^{(m)}_k(\mathbf{n}) = \frac{1}{\npartcl} \sum^{\lattsize}_{\kp=1} \occup_{\kp} \sin\left[2\pi\Eorder \left(\frac{\kp - k}{\lattsize}\right)\right] \, .
\end{equation}
This field is thus reminiscent of a generalized Kuramoto-like long-range interaction term, stressing the mathematical link between the TCLG and the Kuramoto model of oscillator synchronization \cite{kuramoto84a, kuramoto87a, pikovsky03a, acebron05a}. It is important to bear in mind however that this connection is only formal, as the Kuramoto model for oscillators lacks any particle transport in real space, while we are dealing with a $1d$ driven fluid. An additional key difference is the presence of particle exclusion interactions in the TCLG, which becomes apparent in its hydrodynamic transport coefficients and play an important role to understand the time-crystal phase transition, while it is absent in synchronization models.

The evolution of the TCLG model here introduced is described at the microscopic level by a Markovian master equation for the probability of a configuration $\mathbf{n}$ at time $t$ \cite{hurtado-gutierrez20a,hurtado-gutierrez25a}. In the macroscopic limit $\lattsize \to \infty$, using a local equilibrium ansatz and performing a diffusive scaling of space ($x=k/L\in [0,1]$) and time ($t=\tau/L^2$, with $\tau$ being the microscopic time), one can show that the model is characterized by a density field $\densxt$, with $0\le \rho\le 1$, which evolves according to the usual hydrodynamic equation in one-dimensional driven-diffusive systems \cite{spohn12a, derrida07a, bertini15a, lasanta15a, gutierrez-ariza19a}
\begin{equation}
\partial_t \rho = - \partial_x \Big[-D(\rho) \partial_x \rho + \sigma(\rho) E_x[\rho] \Big]
\label{eq:hydro}
\end{equation}
with periodic boundary conditions. In the TCLG model, the diffusion coefficient is constant, $\diffcoef(\dens)=1/2$, and the mobility is given by $\mobility(\dens) = \dens(1 - \dens)$. As expected, the external field depends on the whole density profile and takes the form
$E_x[\rho] = \epsilon +  \lambda \mathcal{E}_x^{(m)}[\rho]$, with a constant driving field $\epsilon$ and a macroscopic packing field now given by $\mathcal{E}_x^{(m)}[\rho] =  \abs{z_m} \sin(\varphi_m - 2\pi\pos \Eorder)$ and written in terms of the macroscopic $m$th-order packing order parameter 
\begin{equation}
z_m[\rho] = \frac{1}{\rho_0} \int_0^1 dx \rho(x,t) \textrm{e}^{\textrm{i}2\pi m x} \equiv \abs{z_m} \textrm{e}^{\textrm{i} \varphi_m}\, ,
\label{z}
\end{equation}
where $\meandens=\int_0^1 dx\,  \rho(x,t)$ is the (constant) global density of the system. 

Interestingly, the homogeneous density profile $\densxt = \meandens$ is always a solution of Eq.~\eqref{eq:hydro}. Perturbing this homogeneous profile and studying the linear stability of the solution under this perturbation, we can determine the critical threshold $\Epampl_c^{(\Eorder)}$ for the time-crystal transition to occur \cite{hurtado-gutierrez20a,hurtado-gutierrez25a},
\begin{equation}
\lambda_c^{(\Eorder)} = 4\pi\Eorder \frac{\diffcoef(\meandens)\meandens}{\mobility(\meandens)}
= \frac{2\pi\Eorder}{1 - \meandens} \, .
\label{eq:critical_point}
\end{equation}
In this way we expect the homogeneous density solution $\densxt = \meandens$ to become unstable for $\lambda>\lambda_c^{(\Eorder)}$, leading to a density field solution with more complex spatio-temporal structure. This will be confirmed in Sec.~\ref{sec:numerics} below through Monte Carlo simulations of the stochastic microscopic model. 

The form of the resulting perturbation beyond the instability is compatible with the emergence of $m$ traveling-wave condensates, $\densxt =\rho_m(\omega_m t - 2\pi mx)$, moving periodically with an angular velocity $\omega_m$. This instability breaks spontaneously the time-translation symmetry of the homogeneous solution, thus giving rise to a continuous time crystal \cite{wilczek12a, shapere12a,zakrzewski12a,richerme17a,yao18a, sacha18a,sacha20a,hurtado-gutierrez20a,hurtado-gutierrez25a}. Note that the value of $\lambda_c^{(\Eorder)}$ increases with $m$, see Eq.~\eqref{eq:critical_point}, meaning that a stronger packing field is needed as $m$ increases to compete against diffusion, which tends to destroy the $m$ emergent condensates \cite{hurtado-gutierrez25a}. Right above the instability ($\lambda\gtrsim \lambda_c^{(\Eorder)}$), the condensates' velocity can be shown to be $\omega_m= 2\pi m \sigma'(\rho_0) \epsilon$ \cite{hurtado-gutierrez25a}. For the TCLG $\sigma'(\rho_0)=1-2\rho_0$, and the sign of the condensate velocity $\omega_m$ thus depends on the initial global density and changes across $\rho_0=1/2$ due to particle-hole symmetry of the model: a particle condensate moving anti-clockwise for $\rho_0>1/2$ can be seen as a hole condensate moving clockwise, and viceversa.
    
On the other hand, using the current field $j(x,t)\equiv -D(\rho) \partial_x \rho + \sigma(\rho) E_x[\rho]$, see Eq.~\eqref{eq:hydro}, the excess of the average current $J=\tau^{-1}\int_0^\tau dt \int_0^1 dx~j(x,t)$ with respect to the homogeneous-phase average current $J_0=\sigma(\rho_0) \epsilon$ can be shown to be $J-J_0\propto \sigma''(\rho_0) \epsilon$ right after the instability \cite{hurtado-gutierrez25a}. In this way the particle current will be larger or smaller than the homogeneous-phase flow depending on the mobility curvature for density $\rho_0$. This highlights the relevance of transport coefficients in the system's response to the packing field, which enhances or lowers the current and the wave velocity depending on the mobility derivatives. For the particular case of the TCLG, where $\sigma''(\rho_0)=-2<0$, we thus expect a negative excess current ($J<J_0$), compatible with the emergence of condensates that jam particle dynamics \cite{perez-espigares13a,hurtado-gutierrez20a,hurtado-gutierrez25a}.

\begin{figure*}
\includegraphics[width=\linewidth]{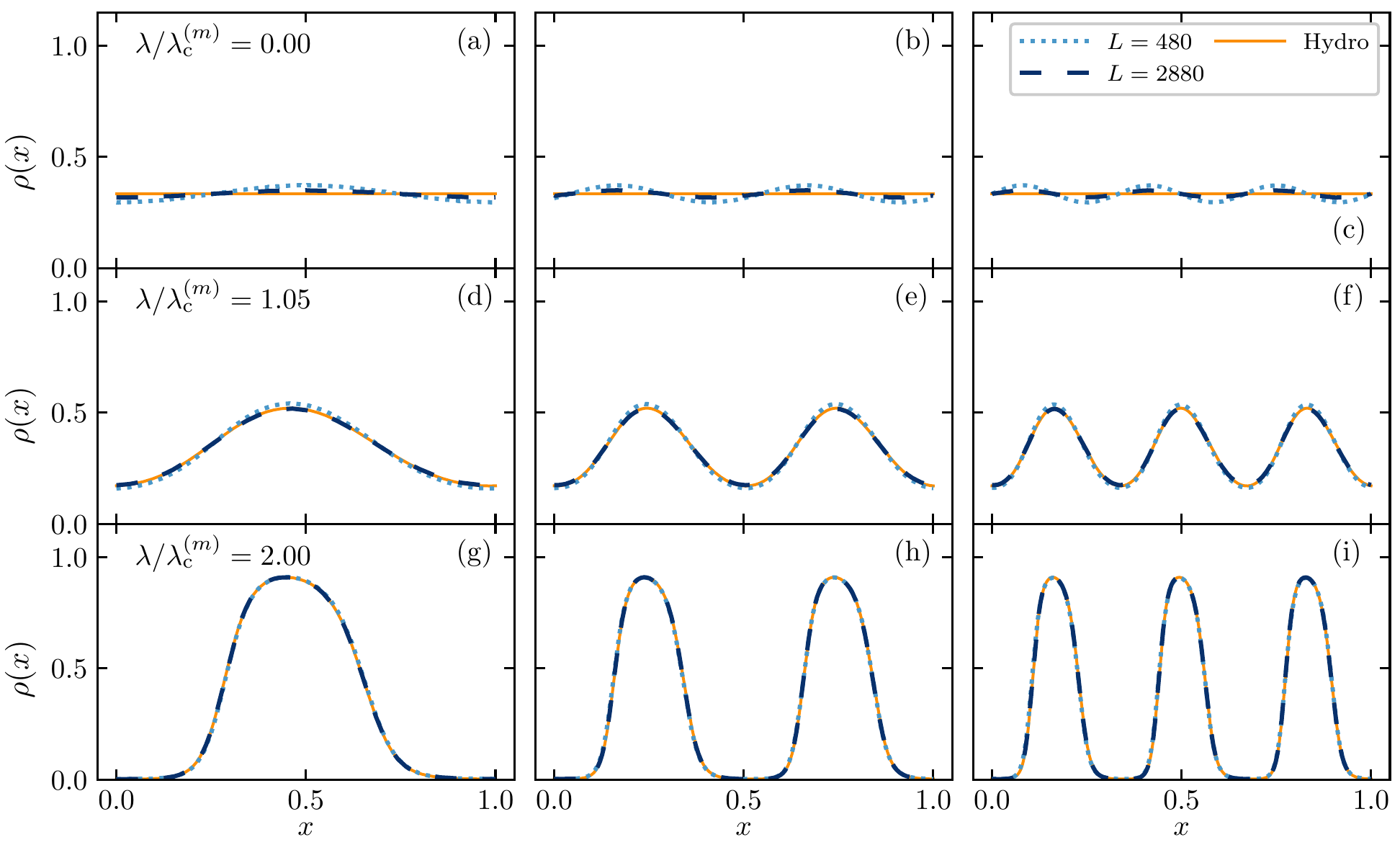}
\caption{\textbf{Density field across the phase transition.}
Measured density profiles for packing orders $\Eorder=1$ (left column), $\Eorder=2$ (central column) and $\Eorder=3$ (right column), as obtained from Monte Carlo simulations of the TCLG for $\rho_0=1/3$, driving field $\Ed = 2.5 \Eorder$, and different values of $L$ (see legend). Panels (a)-(c) show data obtained for packing field coupling $\lambda=0$, corresponding to the homogeneous density phase, panels (d)-(f) display measurements right above the critical point $\lambda_c^{(m)}$, and panels (g)-(i) show data deep inside the time-crystal phase. Full lines correspond to hydrodynamic predictions in all cases, showing an excellent agreement.
}
\label{fig:density_profiles_phases}
\end{figure*}

Finally, we can prove a remarkable scaling property of the emergent custom time-crystal phase under the traveling-wave solution ansatz \cite{hurtado-gutierrez25a}. In particular, it can be shown that the $m$-th-order traveling-wave solution of Eq.~\eqref{eq:hydro} once time-translation symmetry is spontaneously broken can be built by \emph{gluing} together $m$ copies of the $m=1$ solution after an appropriate renormalization of driving parameters, i.e. $\rho_m(\omega_m t - 2\pi mx) = \rho_1(m\omega_1 t - 2\pi mx)$. Here $\rho_1(\omega_1 t - 2\pi x)$ is a traveling-wave solution of Eq.~\eqref{eq:hydro} of velocity $\omega_1$ for $m=1$ and generic driving parameters $\epsilon_1$ and $\lambda_1$, while $\rho_m(\omega_m t - 2\pi mx)$ is the corresponding traveling-wave solution of Eq.~\eqref{eq:hydro} of velocity $\omega_m=m\omega_1$ for arbitrary $m>1$ and driving parameters $\epsilon_m=m\epsilon_1$ and $\lambda_m=m\lambda_1$. This scaling property, valid for arbitrary  nonlinear transport coefficients \cite{hurtado-gutierrez25a}, enables us to collapse travelling-wave solutions for different orders $m$ and related driving parameters, reducing the range of possible solutions. We will check this scaling property in numerical simulations of the microscopic model below.

Note that the Kuramoto synchronization model exhibits a similar equivalence at the hydrodynamic level between first-order and higher-order couplings \cite{delabays19a}, which can be also extended to  dynamics. This \emph{dynamical} equivalence cannot be generalized however to the TCLG model due to the nonlinear character of the mobility transport coefficient, related to the underlying particle exclusion interaction.

\section{Monte Carlo simulations of the time crystal lattice gas}
\label{sec:numerics}
We proceed now to verify the forecasted instability and the different hydrodynamic predictions in Monte Carlo simulations of the microscopic TCLG model introduced in Section \S\ref{sec:model}, comparing with the theoretical results obtained from the numerical integration of the hydrodynamic evolution equation~\eqref{eq:hydro} whenever possible. In particular, we performed standard discrete-time random-sequential Monte Carlo simulations \cite{bortz75a, gillespie77a, newman99a, binder10a}  of the TCLG, where at each step a particle is chosen at random and attempts a biased hop to the right with probability $p_k^+(\mathbf{n})/(p_k^+(\mathbf{n})+p_k^-(\mathbf{n}))$ or to the left with probability $p_k^-(\mathbf{n})/(p_k^+(\mathbf{n})+p_k^-(\mathbf{n}))$ taking into account the exclusion rule. These simulations were carried out at a constant global density $\rho_0=1/3$ and varying values of $L\in[480,2880]$, so as to perform a detailed finite-size scaling analysis of the phase transition. Unless otherwise specified, we will work with a fixed driving field $\Ed_m = 2.5 \Eorder$ for each packing order $\Eorder$. This scaling of the driving field with $m$ will allow us to compare condensate profiles for different values of $m$, as predicted by the hydrodynamic theory. We will however permit variations in $\Ed_m$ later on in this section, in order to explore the effect of the driving field on the resulting particle condensates.
    
\begin{figure*}
\includegraphics[width=\linewidth]{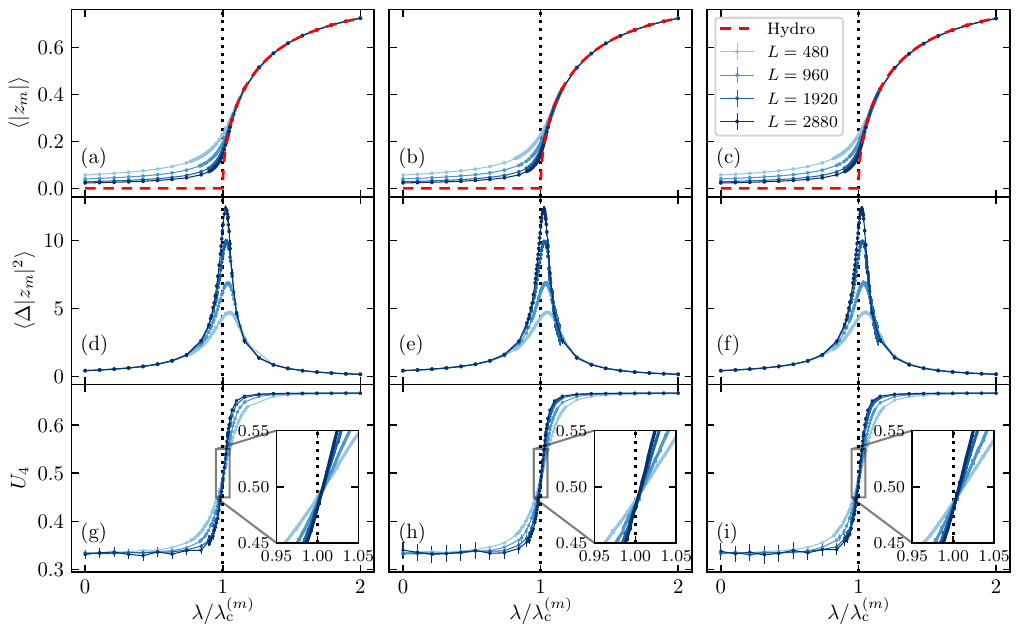}
\caption{\textbf{Characterizing the phase transition.}
Monte Carlo results for the order parameter $\langle |\opar_{\Eorder}|\rangle$ [panels (a)-(c)], its scaled fluctuations $\langle \Delta|z_m|^2\rangle=L (\langle |\opar_{\Eorder}|^2\rangle - \langle |\opar_{\Eorder}|\rangle^2)$ [panels (d)-(f)], and the Binder cumulant $U_4$ [panels (g)-(i)], for the TCLG model as a function of the packing field coupling $\lambda$ and measured for packing orders $\Eorder=1$ (left column), $\Eorder=2$ (central column), and $\Eorder=3$ (right column), global density $\rho_0=1/3$, driving field $\Ed_\Eorder = 2.5 \Eorder$, and different values of $L$ (see legend). The red dashed lines in panels (a)-(c) correspond to the hydrodynamic theory prediction for $\langle |\opar_{\Eorder}|\rangle (\lambda)$, while the vertical dashed lines in all panels signal the predicted critical point $\lambda_c^{(m)}$, see Eq.~\eqref{eq:critical_point}. The solid curves correspond to steady-state time averages of the microscopic simulations. 
}
\label{fig:critical_point_sim}
\end{figure*}

The phase transition is most evident at the configurational level, so we measured the average particle density profiles along the $1d$ ring for different values of the packing field coupling $\lambda$ across the predicted critical point $\lambda_c^{(m)}$, see Eq.~\eqref{eq:critical_point}, different packing orders $m=1,~2,~3$, and varying $L$, see Fig.~\ref{fig:density_profiles_phases}. Due to the system periodicity, and in order not to blur away the possible structure of the density field, we performed profile averages only after shifting the argument of the instantaneous packing order parameter to $x= 1/2$. Note that this procedure leads to a spurious weak structure even in the homogeneous density phase ($\lambda<\lambda_c^{(m)}$) due to the fluctuations of the particles' packing \cite{hurtado11a}, see Figs.~\ref{fig:density_profiles_phases}.(a)-(c), equivalent to averaging noisy homogeneous density profiles around their (random) center of mass. On the other hand, supercritical ($\lambda>\lambda_c^{(m)}$) density fields exhibit a much more pronounced structure resulting from the appearance of traveling particle condensates. In this way, Figs.~\ref{fig:density_profiles_phases}.(a)-(c) thus confirm that before the critical point ($\lambda<\lambda_c^{(m)}$), density profiles remain homogeneous despite the system sustains a net particle current in the direction of the driving field. However, as the packing field coupling $\lambda$ is increased beyond the predicted critical value $\lambda_c^{(m)}$, see Eq.~\eqref{eq:critical_point}, a collection of $m$ equivalent particle condensates emerges. These condensates travel along the $1d$ lattice ring with an approximately constant speed, with weak velocity fluctuations vanishing in the $\lattsize \to \infty$ limit. The emergent matter waves become more prominent and compact as the coupling $\lambda$ increases, see Figs.~\ref{fig:density_profiles_phases}.(g)-(i). Interestingly, the density profiles of the condensates are slightly asymmetric in shape, and this asymmetry increases and becomes more noticeable as $\Epampl$ grows. We will further discuss this asymmetry below, and the role of the constant driving field in its control. Notice also that, in all cases studied, there is a remarkable agreement between the predictions of the hydrodynamic theory for the density profile and the simulation results of the TCLG, see Fig.~\ref{fig:density_profiles_phases}, even for relatively small lattice sizes.

The numerical solution of the integro-differential equation \eqref{eq:hydro} under the traveling-wave ansatz is challenging despite the simplifying character of this assumption, as the system periodicity leads to a two-point boundary value problem which depends on the integrals of the solution. To overcome this issue, we start from Eq.~\eqref{eq:hydro} and use a traveling-wave ansatz $\densxt =\rho_m(\omega_m t - 2\pi mx)$ to obtain an ordinary differential equation (ODE) for $\rho_m(y)$, with $y=\omega_m t - 2\pi mx$, that can be integrated once to arrive at the following first-order ODE \cite{hurtado-gutierrez25a},
\newcommand{\odeconst}{C}
\begin{eqnarray}
\omega_m \rho_m(y) &=& 2\pi \Big\{ D(\rho_m) 2\pi \rho_m^{\prime}(y) \\
&+& \mobility(\rho_m) \big[\Ed + \Epampl \abs{\oparm} \sin(\Eorder y)\big]\Big\} + \odeconst \nonumber
\end{eqnarray}
with $C$ an integration constant, and where we have chosen $\varphi_m(t=0) = 0$ without loss of generality. Taking $|\oparm|$ as a free parameter, the previous equation then turns into a standard ODE with parameters $|\oparm|$, $\omega_m$, and $\odeconst$. This equation can be solved self-consistently using initial ansatzs for $|\oparm|$, $\rho_m(0)$, $\omega_m$ and $\odeconst$, which can be then refined and improved until convergence is achieved taking into account the system periodicity $\rho_m(0) = \rho_m(2\pi)$, and the integral constraints $\meandens = \int_0^{1} \dd x \rho_m(-\wnumber x)$ and $\oparm=\int_0^{1} \dd x \rho_m(-\wnumber x) \textrm{e}^{\textrm{i} 2\pi x \Eorder}$. In practice the convergence of the numerical solution was faster when (i) the initial condition for the density field was obtained from a rough simulation of the full hydrodynamic equation \eqref{eq:hydro}, and (ii) for each proposal for $|\oparm|$, $\rho_m(0)$ and $\omega_m$, we chose $\odeconst$ independently to fix the boundary conditions. This approach is indeed similar to the shooting method \cite{press07a} used to solve two-point boundary value problems. See the appendices in \cite{hurtado-gutierrez25a} for more details.
 
\begin{figure}
\includegraphics[width=\linewidth]{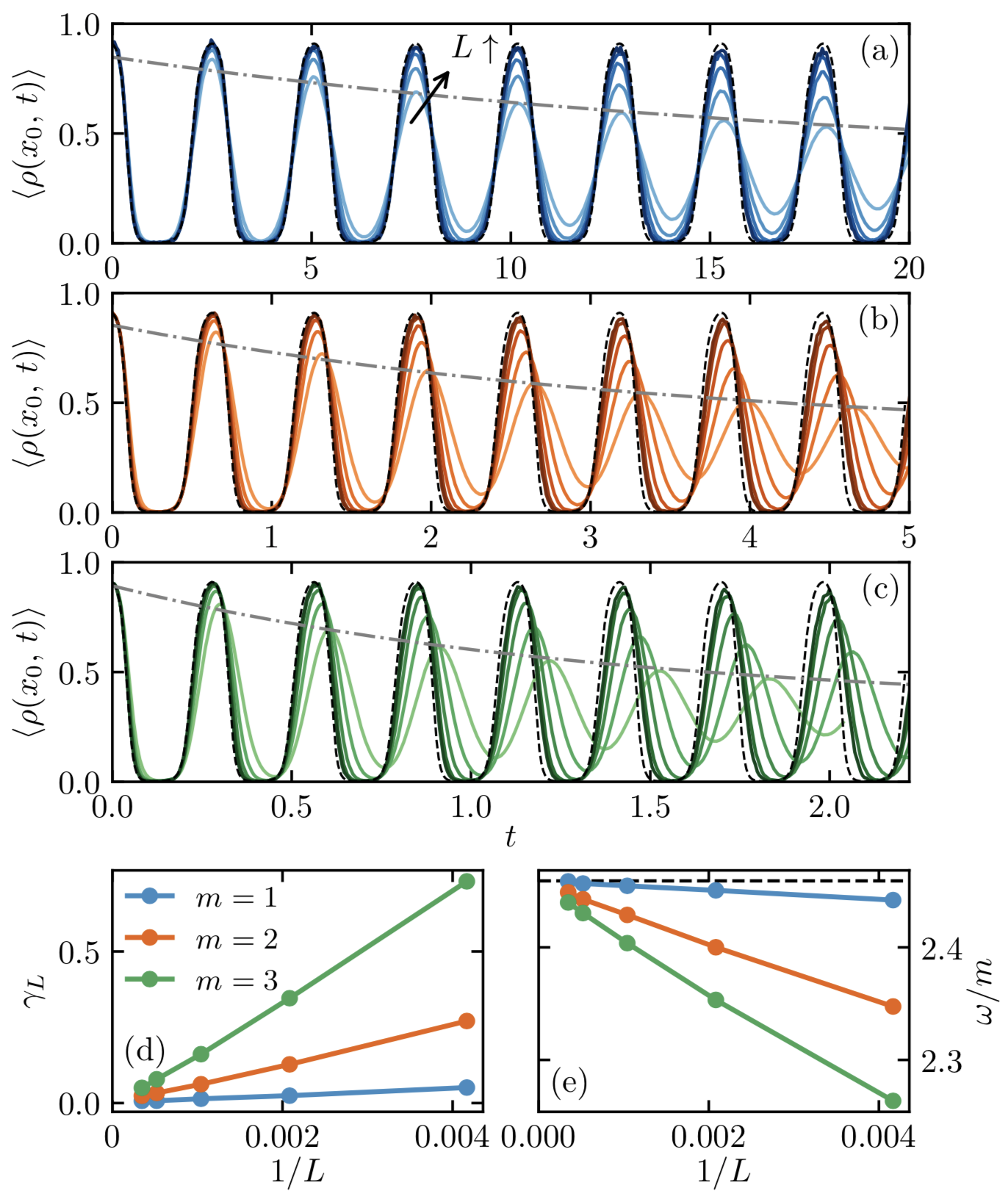}
\caption{\textbf{Rigidity of the time-crystal phase.} 
Panels (a)-(c) show the oscillations of the average density at a given lattice site, $\langle \rho(x_0,t)\rangle$, for packing orders $m = 1,~2,~3$, respectively, parameters $\epsilon_m = 2.5m$ and $\lambda = 2\lambda_c^{(m)}$, and for different lattice sizes $L = 240,~480,~960,~1920,~2880$ (plotted in increasing color intensity). Note that the amplitude of oscillations is damped in time for finite lattice sizes. The black dashed line shows the hydrodynamic solution in each case, while the gray dash-dotted line displays the exponential envelope of the $L = 240$ curves. Panel (d) displays the measured damping rate $\gamma_L$ of the oscillations as a function of $1/L$, showing how it vanishes in the macroscopic limit, a signature of the rigidity of the resulting time-crystal phases in this limit. Panel (e) illustrates the convergence of the measured (finite-size) angular speed $\omega_m$ of the traveling wave to the hydrodynamic prediction, shown as a black horizontal dashed line. In all cases, error bars are not shown because they are smaller than the markers. For all choices of the parameters, results were obtained from averaging at least 1000 runs of the simulations.
}
\label{fig:oscillations}
\end{figure}
    
\begin{figure*}
\includegraphics[width=\linewidth]{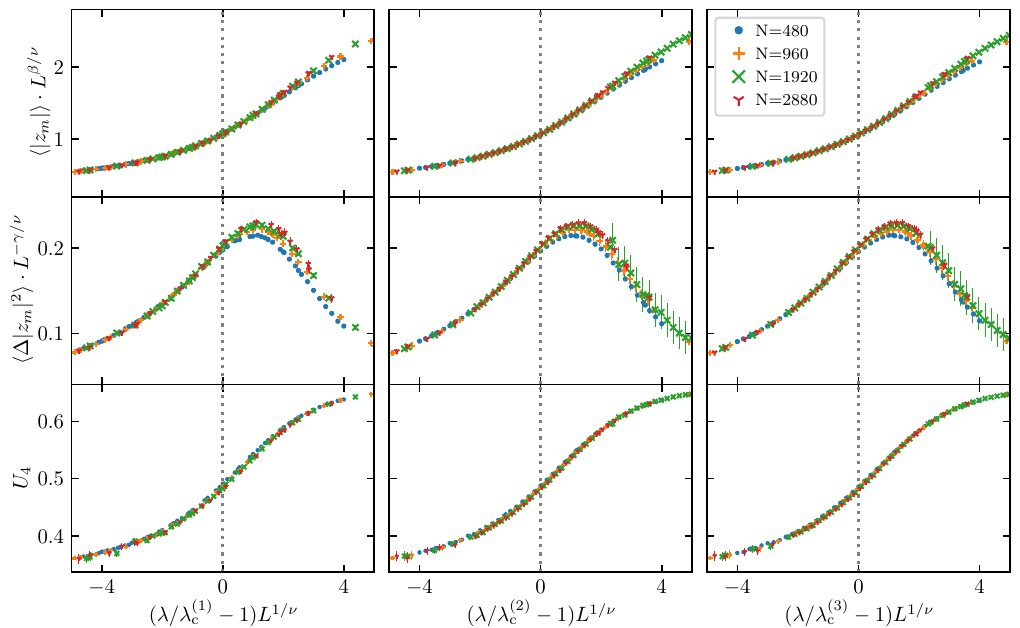}
\caption{\textbf{Data collapses around the critical region.} Scaling plots for the order parameter $\langle |\opar_{\Eorder}|\rangle$ (top row), the susceptibility $\langle \Delta|z_m|^2\rangle$ (middle row), and the Binder cumulant $U_4$ (bottom row), using all data shown in Fig.~\ref{fig:critical_point_sim} (corresponding to $\rho_0=1/3$, $L\in[480,2880]$, $m=1,~2,~3$ and $\Ed(\Eorder) = 2.5 \Eorder$) and the critical exponents $\nu=2$, $\gamma=1$, $\beta=1/2$ corresponding to the (mean field) Kuramoto universality class. Vertical dashed lines signal the critical point.
}
\label{fig:critical_exponents_collapse}
\end{figure*}

As discussed in the previous section, a good order parameter for the reported phase transition is the average magnitude of the packing order parameter, $\langle |\opar_{\Eorder}|\rangle$, which is a measure of the particles' condensation around the $m$ emergent localization centers. Figs.~\ref{fig:critical_point_sim}.(a)-(c) show the measured $\langle |\opar_{\Eorder}|\rangle$ as a function of the packing field coupling $\lambda$ for $\rho_0=1/3$, $m=1,~2,~3$ and different system sizes $L$. As expected, the order parameter remains small for $\lambda<\lambda_c^{(m)}$, approaching zero in the thermodynamic $L\to\infty$ limit. However, upon crossing $\lambda_c^{(m)}$, the value of $\langle |\opar_{\Eorder}|\rangle$ increases sharply but in a continuous way, a behavior typical of second-order continuous phase transitions. Remarkably, the measured $\langle |\opar_{\Eorder}|\rangle(\lambda)$ agrees nicely with the hydrodynamic predictions for the order parameter $\forall m$ even for moderate system sizes, see red dashed line in Figs.~\ref{fig:critical_point_sim}(a)-(c). We also measured the susceptibility across the transition as captured by the order parameter scaled fluctuations, $\langle \Delta|z_m|^2\rangle \equiv L\left(\langle |\opar_{\Eorder}|^2\rangle - \langle |\opar_{\Eorder}|\rangle^2\right)$,  see Figs.~\ref{fig:critical_point_sim}.(d)-(f). The susceptibility $\langle \Delta|z_m|^2\rangle$ exhibits a steep peak around $\lambda_c^{(m)}$, the sharper the larger the system size $L$, strongly suggesting a divergence of the scaled order parameter fluctuations at the critical point $\lambda_c^{(m)}$ in the thermodynamic limit, as expected for a continuous phase transition. Note that, for each $\lattsize$, the finite-size peak in $\langle \Delta|z_m|^2\rangle$ is always located slightly after the critical point, $\lambda^* \gtrsim \lambda_c^{(m)}$, an observation in contrast with the behavior reported in the Kuramoto model of synchronization \cite{daido90a, hong15a} and most probably related with the particle exclusion interactions that characterize the TCLG model. In order to determine the critical packing field coupling $\lambda_c^{(m)}$ from simulations, we also measured the Binder cumulant $\binder=1 - \avg{|\opar_{\Eorder}|^4}/(3\avg{|\opar_{\Eorder}|^2}^2)$ as a function of $\lambda$ \cite{binder10a}, see Figs.~\ref{fig:critical_point_sim}(g)-(i). It is well known that, due to the finite-size scaling properties of this observable, the curves $U_4(\lambda)$ measured for different values of $L$ should cross at an $L$-independent value of $\lambda$ that identifies the infinite-size critical point $\lambda_c^{(m)}$. This is indeed clearly observed in Figs.~\ref{fig:critical_point_sim}.(g)-(i), where we can also verify that these measured critical packing field couplings agree remarkably well with the hydrodynamic predictions for $\lambda_c^{(m)}$ $\forall m$ (vertical dashed lines), obtained from a local stability analysis of the homogeneous density solution, see Eq.~\eqref{eq:critical_point}.

Interestingly, if we monitor the average density at a given point along the ring, e.g. $\langle \rho(x_0,t)\rangle$, we find persistent oscillations as a function of time for any packing order $m$, see Fig.~\ref{fig:oscillations}, a universal signature of time-crystalline order \cite{wilczek12a, shapere12a, zakrzewski12a, bruno13a, nozieres13a, watanabe15a, khemani16a, else16a, keyserlingk16a, yao17a, moessner17a, richerme17a, zhang17a, choi17a, nakatsugawa17a, lazarides17a, sacha18a, gambetta19a, gong18a, tucker18a, osullivan18a, yao18a, yao20a, iemini18a, gambetta19b, medenjak20a, lazarides20a, sacha20a}. These oscillations are characterized by a period $2\pi\omega_m^{-1}$ in the diffusive scale, with $\omega_m$ the condensates average velocity. However, for any finite system size $L$, the oscillations amplitude is clearly damped in time, see Fig.~\ref{fig:oscillations}, with a damping rate $\gamma_L$ that can be obtained from an exponential fit to the envelope of $\langle \rho(x_0,t)\rangle$ for each $L$, see dot-dashed lines in Figs.~\ref{fig:oscillations}.(a)-(c). In any case, the measured $\gamma_L$ decays to zero for each $m$ in the thermodynamic $L\to \infty$ limit, see Fig.~\ref{fig:oscillations}.(d), meaning that the observed local density oscillations converge toward the hydrodynamic (undamped) periodic prediction in this limit (dashed line in Figs.~\ref{fig:oscillations}.(a)-(c)). This observation clearly signals the rigidity of the long-range spatiotemporal order emerging in the observed programmable time-crystal phases. We also measured the condensate speed $\omega_m$ for different packing orders $m$ as a function of $1/L$, see Fig.~\ref{fig:oscillations}.(e), finding a remarkable agreement with the hydrodynamic predictions in the large system size limit $L\to \infty$.

In order to determine the critical exponents characterizing the universality class of the observed phase transitions to these programmable time-crystal phases, we performed extensive measurements around the critical region in our Monte Carlo simulations. Due to the scale invariance underlying any continuous phase transitions, the relevant observables near the critical point are expected to be homogeneous functions of the two length scales characterizing the problem at hand, namely the correlation length $\xi$ and the system size $L$ \cite{binney92a, binder10a}. This immediately implies the following scaling relations near the critical point,
\begin{eqnarray}
|\opar_{\Eorder}|(\Epampl, \lattsize) &=& \lattsize^{-\oparexp/\clengthexp} \mathcal{F}\left[(1-\Epampl/\lambda_c^{(m)})\lattsize^{1/\clengthexp}\right] \, , \nonumber \\
\Delta |\opar_{\Eorder}|^2(\Epampl, \lattsize) &=& \lattsize^{\suscepexp/\clengthexp} \mathcal{G}\left[(1-\Epampl/\lambda_c^{(m)})\lattsize^{1/\clengthexp}\right] \, , \nonumber \\
\binder(\Epampl, \lattsize) &=& \mathcal{H}\left[(1-\Epampl/\lambda_c^{(m)})\lattsize^{1/\clengthexp}\right] \, , \nonumber
\end{eqnarray}
where $\oparexp$, $\suscepexp$ and $\clengthexp$ are the critical exponents characterizing the power-law behavior of the order parameter, the susceptibility and the correlation length, respectively. The functions $\mathcal{F}(\zeta)$, $\mathcal{G}(\zeta)$ and $\mathcal{H}(\zeta)$ are the corresponding master curves or scaling functions characterizing the underlying universality class. These scaling relations are tested for the TCLG model in Fig.~\ref{fig:critical_exponents_collapse}. In particular, the best data collapses for the order parameter $\langle |\oparm|\rangle$, the susceptibility $\langle \Delta|z_m|^2\rangle$, and the Binder cumulant $\binder$ are obtained when using the mean-field exponents $\oparexp = 1/2$, $\suscepexp$ = 1, and $\clengthexp=2$. These exponents characterize the universality class of the Kuramoto synchronization transition \cite{pikovsky03a}. All scaling plots display a remarkable collapse for different lattice sizes. Note however that, while the collapse for the susceptibility is excellent for $\lambda<\lambda_c^{(m)}$, it falls off slightly for $\lambda>\lambda_c^{(m)}$, an asymmetric scaling behavior reminiscent of that already observed in some models of synchronization \cite{hong15a}. Interestingly, our results thus strongly suggest that the universal scaling properties around the transition to these programmable time-crystal phases (which breaks continuous time-translation symmetry) do not depend on the number $m$ of emergent particle condensates induced by the packing field or other details of the driving mechanism. Indeed the scaling curves for different $m$ also collapse on top of each other, see Fig.~\ref{fig:critical_exponents_collapse}.

\begin{figure}
\includegraphics[width=\linewidth]{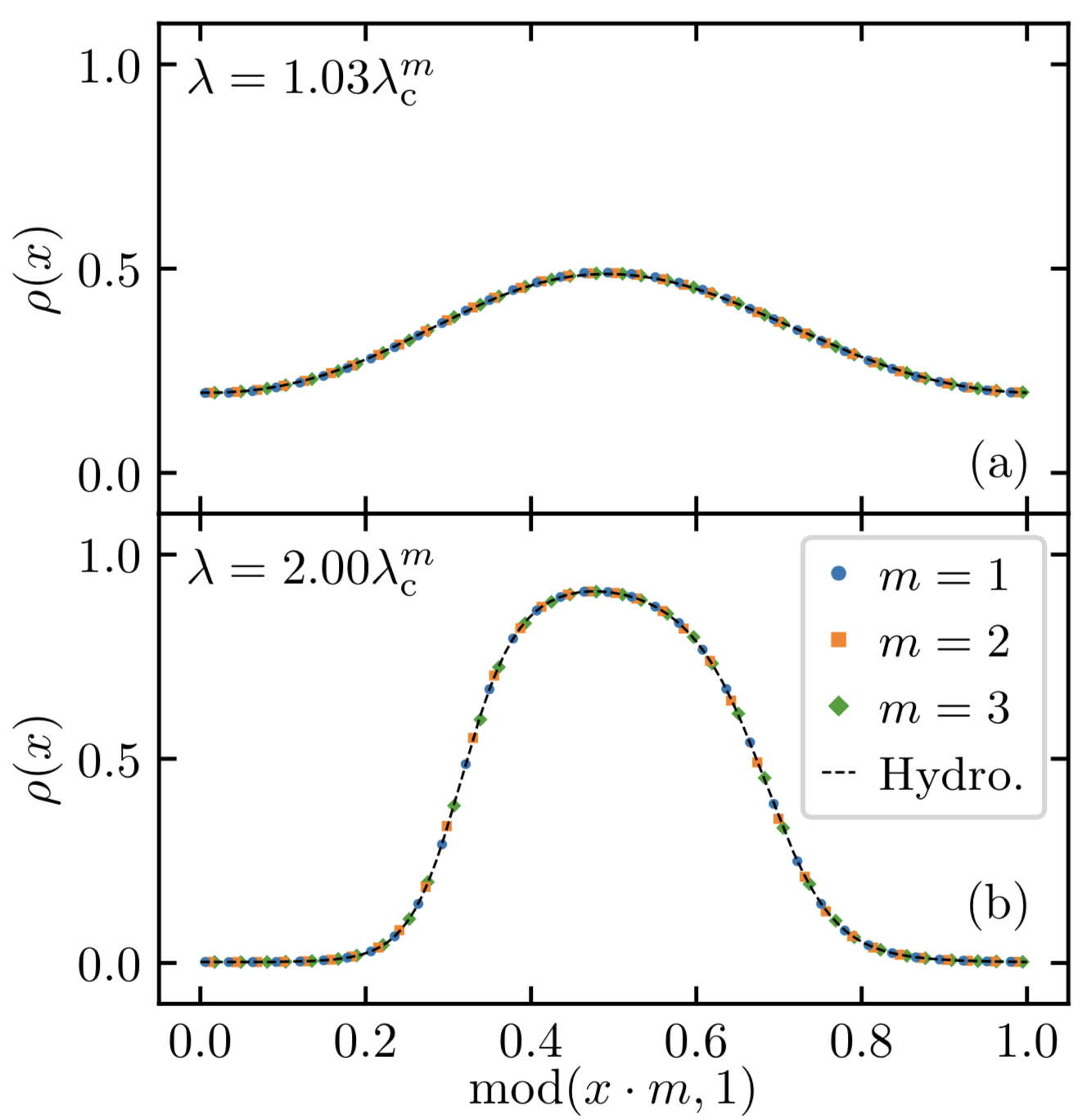}
\caption{\textbf{Invariance of condensate profiles.}
Collapse of the measured condensate density profiles for different packing orders $\Eorder = 1,~2,~3$, for $\lambda=1.03\lambda_c^{(m)}$ (top panel) and $\lambda=2\lambda_c^{(m)}$ (bottom panel). The different markers correspond to simulation results with $\lattsize = 1920$ and various $m$, while the dashed line corresponds to the scaled hydrodynamic solution. The driving field for each packing order is $\Ed = 2.5\Eorder$, and the global density is $\rho_0=1/3$.
}
\label{fig:density_profiles_collapse}
\end{figure}

The question remains as to whether the shape of the emergent particle condensates in the time-crystal phase is also independent of the packing order $m$ after some renormalization of parameters, as it is indeed predicted by hydrodynamics, see Eqs.~\eqref{eq:hydro}-\eqref{eq:critical_point}. In particular, the prediction is that the $m$-th-order traveling-wave solution of Eq.~\eqref{eq:hydro} once time-translation symmetry is spontaneously broken corresponds to $m$ copies of the $m=1$ solution after an appropriate renormalization of driving parameters. This means that $\rho_m(\omega_m t - 2\pi mx) = \rho_1(m\omega_1 t - 2\pi mx)$, where $\rho_1(\omega_1 t - 2\pi x)$ is a traveling-wave solution of Eq.~\eqref{eq:hydro} of velocity $\omega_1$ for $m=1$ and generic driving parameters $\epsilon_1$ and $\lambda_1$, while $\rho_m(\omega_m t - 2\pi mx)$ is the corresponding traveling-wave solution of Eq.~\eqref{eq:hydro} of velocity $\omega_m=m\omega_1$ for arbitrary $m>1$ and driving parameters $\epsilon_m=m\epsilon_1$ and $\lambda_m=m\lambda_1$. To address this question in simulations, we measured in detail the density profiles for different values of $\lambda>\lambda_c^{(m)}$ deep into the time-crystal phase for $m=1,~2,~3$ and parameters $\Ed$ and $\Epampl$ scaled as described above, see Fig.~\ref{fig:density_profiles_collapse}. In agreement with the hydrodynamic theory, the measured condensate profiles match and collapse neatly on top of each other, $\rho_m(x)=\rho_1(mx)$, once the spatial coordinate $x$ is stretched by the corresponding packing order $\Eorder$. Moreover, the collapsed shape agrees accurately with the hydrodynamic curve predicted by the traveling-wave solution, see dashed lines in Fig.~\ref{fig:density_profiles_collapse}.
        
\begin{figure}
\includegraphics[width=\linewidth]{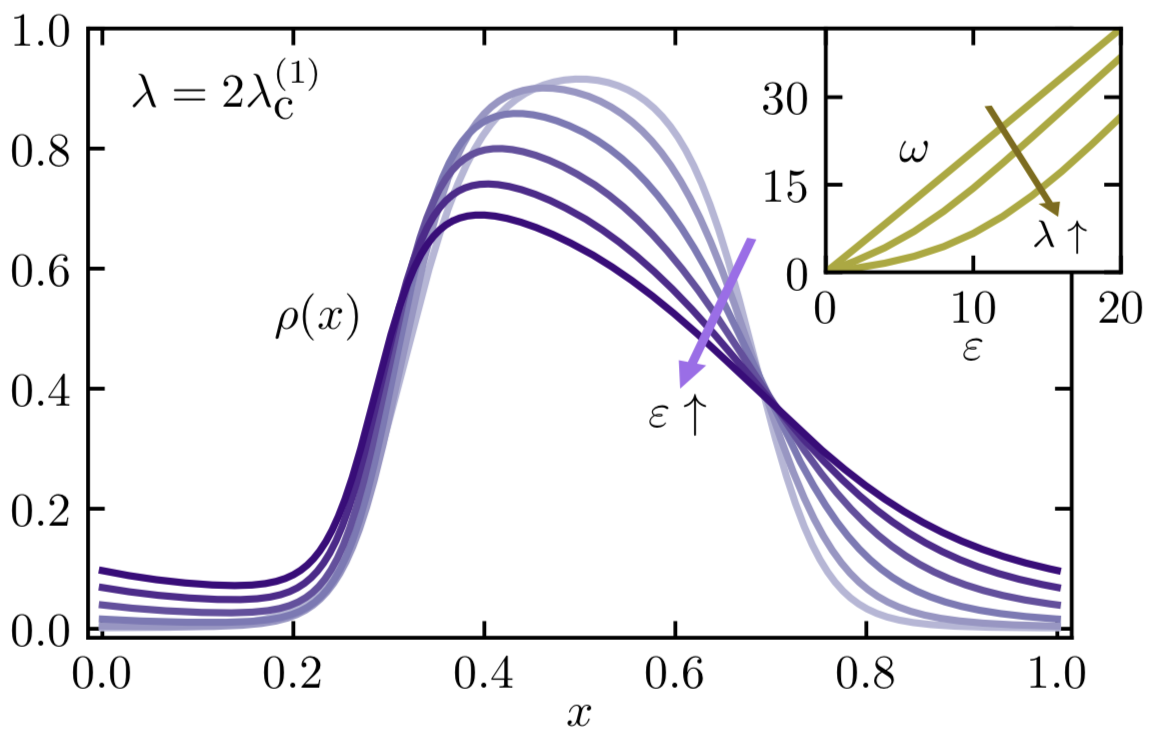}
\caption{\textbf{Condensate asymmetry and velocity.}
Main: Solution of the hydrodynamic traveling-wave equation~\eqref{eq:hydro} for $\Eorder=1$, $\Epampl=2\lambda_c^{(1)}$ and driving fields $\Ed=0, 2, 4, 6, 9, 10$. The asymmetry of the condensate density profile increases as $\Ed$ grows. Inset: Traveling-wave velocity as a function of the driving field $\Ed$ and several supercritical couplings $\lambda$, as obtained from hydrodynamics.
}
\label{fig:twprofile_vs_a}
\end{figure}

An interesting trait of condensate profiles deep into the time-crystal phase is their spatial asymmetry, which reflects the nonlinear transport properties of model. This is already apparent in Figs.~\ref{fig:density_profiles_phases}.(g)-(i) and Fig.~\ref{fig:density_profiles_collapse}.(b). Indeed, for the TCLG the particle current in the time-crystal phase is lower than in the homogeneous phase due to the exclusion interaction (i.e. negative excess current, $J<J_0$, see Section \S\ref{sec:model}), meaning that the emergence of condensates jams dynamics on average. This jamming gives rise in turn to a sharp particle accumulation at the rear tail of the condensate, while the condensate front displays a soft decay. Indeed, the leading front is smeared out by the action of the driving field because density along its line of motion is low and exclusion plays little role. On the other hand, the rear tail motion is strongly hindered by the high density of the particle condensate, thus leading to the observed asymmetric condensate shape. The stronger the driving field, the more important this difference between the leading and rear fronts is, resulting in an increasing asymmetry. This is fully confirmed in Fig.~\ref{fig:twprofile_vs_a}, which shows the condensate density profiles as obtained from the traveling-wave solution of the hydrodynamic theory \eqref{eq:hydro} for $m=1$, $\Epampl>\lambda_c^{(1)}$ and several driving fields $\epsilon$. For completeness, we also explore in the inset to Fig.~\ref{fig:twprofile_vs_a} the effect of the driving field on the condensate velocity, which increases with $\Ed$ as expected. Interestingly, this increase in velocity is superlinear for weak and moderate driving fields ($\Ed\lesssim 10$), becoming progressively linear for $\Ed> 10$.
 
\section{Discussion}

Continuous time crystals are many body systems whose ground state breaks time-translation symmetry, a phenomenon that leads to robust periodic motion and thus rigid long-range order in time \cite{wilczek12a,shapere12a,zakrzewski12a,richerme17a,yao18a,sacha18a}. Recent works \cite{hurtado-gutierrez20a,hurtado-gutierrez25a} have introduced a novel mechanism to engineer programmable time-crystal phases in driven diffusive fluids, based on the concept of packing field. This was inspired by remarkable dynamical phase transitions observed in the rare transport fluctuations of some nonequilibrium systems \cite{bodineau05a, hurtado11a, perez-espigares13a, hurtado14a, tizon-escamilla17b, perez-espigares19a, hurtado-gutierrez23a}. Such packing field excites the $m$th-order Fourier modes of the density field around $m$ emergent localization centers, leading eventually to the spontaneous emergence of time-crystalline order in the form of $m$ rotating high-density condensates, which can be further controlled and modulated.

In this work we have explored at the microscopic level the broad possibilities of the packing-field route to time crystals. For that, we have first analyzed the underlying nonequilibrium phase transition to these custom time-crystal phases in a paradigmatic model of particle diffusion under exclusion interactions, the time-crystal lattice gas, a generalization of the weakly asymmetric simple exclusion process under the action of a configuration-dependent $m$th-order packing field \cite{hurtado-gutierrez20a, hurtado-gutierrez25a}. We have carried out extensive Monte Carlo simulations of the TCLG for different values of the packing-field order $m$, measuring the order parameter in each case, the susceptibility and the Binder cumulant, and comparing with the hydrodynamic predictions whenever possible. A finite-size scaling analysis of the results have allowed us to determine the critical points, which fully agree with those predicted by a hydrodynamic stability analysis, and the critical exponents (as well as the universal scaling functions) which characterize the universality class of these time-crystal phase transitions. These exponents turn out to be compatible with the Kuramoto universality class that characterizes the synchronization of oscillators, independently of the packing order $m$. We also characterize the condensates density profiles and velocities, confirming a scaling property predicted for the higher-order condensate shapes in terms of first-order ones.

\begin{figure}
\includegraphics[width=\linewidth]{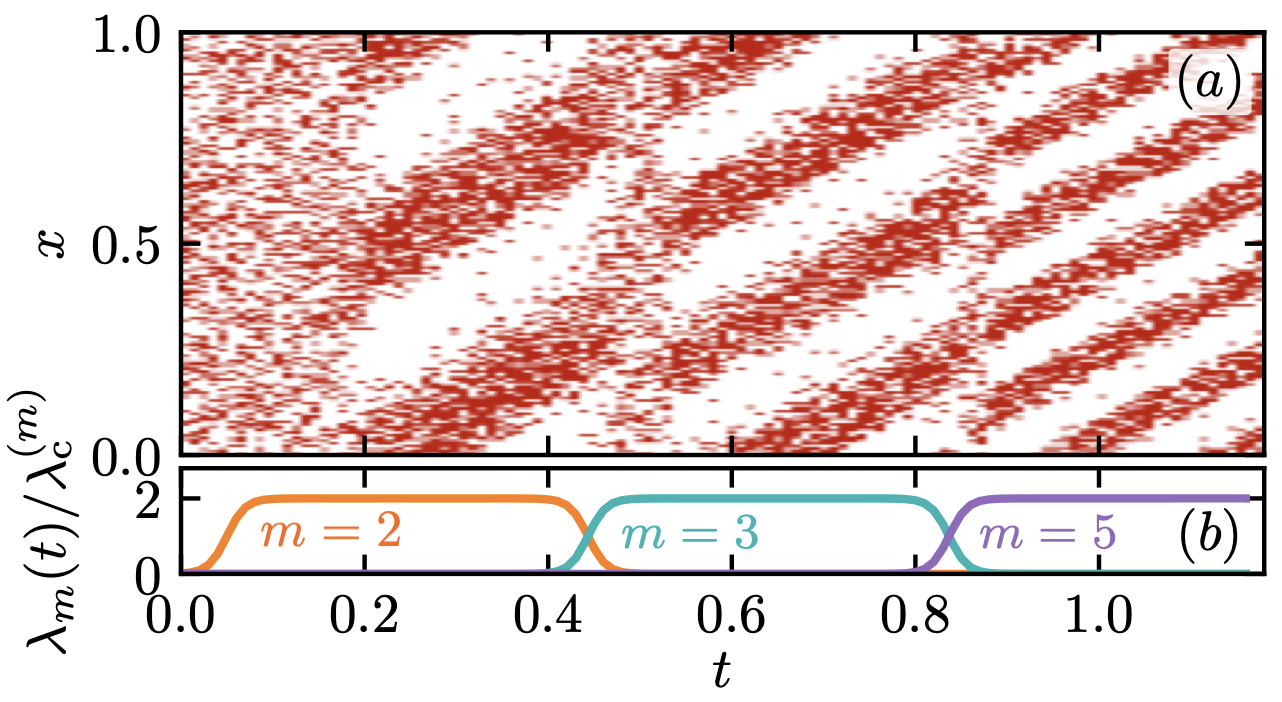}
\caption{\textbf{Engineering programmable time-crystal phases.} (a) Raster plot illustrating the spatiotemporal dynamics of the particle configuration in the TCLG model under the influence of a time-modulated generalized external field $E_{k,t}(\mathbf{n})$, see Eq.~\eqref{eq:genPF}, for $\meandens=1/3$ and $\Ed=0.5$. The system undergoes dynamic transitions between different numbers of condensates as specific orders $m=2,3,5$ are alternately activated or deactivated by modulating $\lambda_m(t)$ beyond the corresponding critical thresholds $\lambda_c^{(m)}$, as depicted in panel (b). This control protocol results in a periodic, intricate pattern decorating the particle distribution over time.
}
\label{fig:raster}
\end{figure}

The packing-field mechanism explored at the microscopic level in this paper is robust and versatile, and allows for the possibility of engineering custom time-crystal phases on demand, with multiple rotating condensates \cite{hurtado-gutierrez20a,hurtado-gutierrez25a}. These time-crystal phases can be further enhanced with higher-order matter waves by introducing competing packing fields modulated in time \cite{hurtado-gutierrez25a}. As a microscopic proof of concept, let us consider a generalized external field
\begin{equation}
E_{k,t}(\mathbf{n}) = \epsilon + \sum_m \lambda_m(t) \mathcal{E}_k^{(m)}(\mathbf{n}) \, ,
\label{eq:genPF}
\end{equation}
where $\lambda_m(t)$ is the time-modulated coupling to the $m$th-order packing field $\mathcal{E}_k^{(m)}(\mathbf{n})$, see also Eq.~\eqref{Epack}.  Fig.~\ref{fig:raster} displays a particular trajectory of the TCLG model under such a time-modulated generalized packing field. In particular, it shows the spatiotemporal evolution of the TCLG microscopic particle configuration subject to a generalized external field which swaps between different number of condensates in time, see Fig.~\ref{fig:raster}.(a), by switching on and off different orders $m$ modulating $\lambda_m(t)$ as in Fig.~\ref{fig:raster}.(b). This is just an example, though one can imagine a myriad of potentially interesting combinations, see \cite{hurtado-gutierrez25a} for other decorated time-crystal phases. This demonstrates the huge potential of the packing-field route to engineer and control custom time-crystal phases in stochastic particle systems, opening new avenues of future research with promising technological applications. The challenge remains to exploit this route to time crystals in more complex geometries and higher dimensions.

Further, the packing-field mechanism explored here could be experimentally tested using colloidal fluids confined in quasi-$1d$ circular optical traps \cite{cereceda-lopez23a,cereceda-lopez23b,cereceda-lopez24a} or channels \cite{lutz04a,villada-balbuena21a}. In particular, optical tweezers \cite{kumar18a,grier97a,ortiz14a,martinez15a,martinez17a,rodrigo18a} could be employed to implement the packing field by dynamically adjusting forces based on the system configuration, via feedback loop protocols. Such experimental setups, already used to study nonequilibrium phase behavior in colloidal systems, offer a promising platform to realize and probe the time-crystal phases discussed in this work.

\begin{acknowledgments}
The research leading to these results has received funding from the I+D+i grants PID2023-149365NB-I00, PID2020-113681GB-I00, PID2021-128970OA-I00, C-EXP-251-UGR23 and  P20\_00173, funded by MICIU/AEI/10.13039/501100011033/, ERDF/EU, and Junta de Andaluc\'{\i}a - Consejer\'{\i}a de Econom\'{\i}a y Conocimiento, as well as from fellowship FPU17/02191 financed by the Spanish Ministerio de Universidades. We are also grateful for the the computing resources and technical support provided by PROTEUS, the supercomputing center of Institute Carlos I in Granada, Spain.
\end{acknowledgments}

\bibliography{/Users/phurtado/PAPERS/BIBLIOGRAPHY/referencias-BibDesk-OK.bib}{}

\end{document}